\documentstyle[aps,epsfig]{revtex}
\begin{document}
\draft
\title {Dynamics of Coupling Functions in Globally Coupled Maps: \\
 Size, Periodicity and Stability of Clusters}
\author{M. G. Cosenza$^*$ and A. Parravano}
\address{Centro de Astrof\'{\i}sica Te\'orica, Facultad de Ciencias,
Universidad de Los Andes, \\ Apartado Postal 26 La~Hechicera, M\'erida~5251,
Venezuela.\\  {\sf [Accepted in Phys. Rev. E (2001)] }}
\maketitle
\begin{abstract}
It is shown how different globally coupled map systems can be analyzed under a common
framework by focusing on the dynamics of their respective global coupling functions.
We investigate how the functional form of the coupling determines the formation of
clusters in a globally coupled map system and the resulting periodicity of the global
interaction. The allowed distributions of elements among periodic clusters is also
found to depend on the functional form of the coupling. Through the analogy between
globally coupled maps and a single driven map, the clustering behavior of the former
systems can be characterized. By using this analogy, the dynamics of periodic clusters
in systems displaying a constant global coupling are predicted; and for a particular
family of coupling functions, it is shown that the stability condition of these
clustered states can straightforwardly be derived.
\end{abstract}
\pacs{PACS Number(s): 05.45.+b, 02.50.-r\\
$^*${\it mcosenza@ciens.ula.ve}}


\section{Introduction}
There has been much interest in the study of the collective dynamics of chaotic
systems subjected to global interactions. Such systems arise naturally in the
description of arrays of Josephson junctions, charge density waves, multimode lasers,
neural dynamics, evolutionary, chemical and social networks
\cite{Hadley,Wiesenfeld,Strogatz,Kuramoto,Kaneko}. The globally coupled map (GCM)
lattice \cite{Kan1} constitutes  a prototype model for such global-coupling dynamics.
It has recently been argued that GCM systems yield universal classes of collective
phenomena \cite{Kanbook}. Specifically, a GCM system can exhibit a variety of
collective behaviors such as clustering (i.e., the formation of differentiated subsets
of synchronized elements in the network) \cite{Kan2}; non-statistical properties in
the fluctuations of the mean field of the ensemble \cite{Kan2}; global quasiperiodic
motion \cite{Kan3,Pikovsky}; and different collective phases depending on the
parameters of the system \cite{Kan3}. It has been shown that a GCM system is closely
related to a single map subjected to an external drive and that this analogy may be
used to describe the emergence of clusters in GCM systems in geometrical terms
\cite{PC}.

In particular, the phenomenon of clustering is relevant as it can provide a simple
mechanism for segregation, ordering and onset of differentiation of elements in many
physical and biological systems. In addition to GCM systems, dynamical clustering has
also been found in a globally coupled R\"{o}ssler oscillators \cite{Mik}, neural
networks \cite{Zan}, and coupled biochemical reactions \cite{Furusawa}. The interest
in this phenomenon has recently grown, since dynamical clusters have been observed
experimentally in an array of electrochemical oscillators interacting through a global
coupling \cite{Kiss}.

In this paper, we investigate the process of cluster formation in general globally
coupled map systems by focusing on the dynamics of their global coupling functions. In
most studies on GCM systems, the mean field of the network has been used as the global
coupling function. Here, we study GCM systems subjected to different global coupling
functions and show how they can be analyzed under a common framework. We investigate
how the distribution of elements among a few clusters and their periodicities depend
on the functional form of the global coupling.

Section II contains a description of the dynamics of different global coupling
functions in GCM systems and a calculation of the possible periodicities and cluster
sizes when two clusters emerge in these systems. The driven map analogy is employed in
Sec. III to interpret the clustering behavior of GCM systems. In Sec. IV the dynamical
properties of periodic clusters in systems exhibiting a constant global coupling are
predicted; and for a particular family of global coupling functions, the stability
condition for these clustered states is derived in an Appendix. Conclusions are
presented in Sec. V.

\section{Dynamics of global coupling functions}
Consider a general globally coupled map system
\begin{equation}\label{gcm}
  x_{t+1}(i)= (1-\epsilon) f\left(x_t(i)\right) + \epsilon
  H(x_t(1),x_t(2),\ldots,x_t(N)),
\end{equation}
where $x_t(i)$ gives the state of the element $i$ $(i=1,2,\ldots,N)$ at discrete time
$t$; $N$ is the size of the system; $\epsilon$ is the coupling parameter; $f(x)$
describes the (nonlinear) local dynamics, which in the present article is chosen to be
the quadratic map $f(x)=1-r x^2$; and $H_t\left(x_t(i),x_t(2),\ldots,x_t(N)\right)$ is
the global coupling function. We shall consider a general class of global coupling
functions of $N$ variables such that
$H(\ldots,x_t(i),\ldots,x_t(j),\ldots)=H(\ldots,x_t(j),\ldots,x_t(i),\ldots)$,
$\forall \, i,j$; that is, $H$ is assumed to be invariant to argument permutations.
This property of the coupling function ensures that, at any time, each element of the
globally coupled system is subjected to the same influence of the coupling term. Some
examples of coupling functions belonging to this class are
\begin{eqnarray}
\label{coupling1}
 H= & \langle x \rangle & =  \frac{1}{N}\sum_{i=1}^N x_t(i); \\
 \label{coupling2}
 H= & \langle f \rangle & =  \frac{1}{N}\sum_{i=1}^N f(x_t(i)); \\
 \label{coupling3}
  H= & \Delta x & =  \sqrt{\frac{1}{N}\sum_{i=1}^N \left( x_t(i)-\langle x \rangle
  \right)^2}; \\
 \label{coupling4}
   H= & \bar{x} & =  \prod_{i=1}^N |x_t(i)|^{\frac{1}{N}}.
\end{eqnarray}

The first two examples correspond to forward and backward mean field coupling,
respectively, and they have been widely used in GCM studies. The third global coupling
function is the usual dispersion or mean square deviation of $N$ variables, and it may
describe systems whose elements do not interact when they are synchronized. This kind
of global interaction might be relevant in some biological or social systems
where the members of a community are driven by their deviations from the mean behavior. 
The last example is the geometric mean. 
This type of multiplicative coupling occurs, for instance, in a system of $N$ sequential 
amplifiers where the gain of element $i$ is a function of the magnitude of its state $x_t(i)$, 
and $H$ is proportional to the total gain of the system.
Many statistical functions of $N$ variables share the
property of invariance under argument permutations and they could as well be taken as
global coupling functions in GCM systems given by Eq.(\ref{gcm}).

For some range of its parameters the GCM system in Eq.(\ref{gcm}) reaches an
asymptotic collective behavior characterized by the segregation of the elements into
$K$ clusters, each exhibiting a period $P$, where the $k \mbox{th}$ cluster has a
number $N_k$ of elements, with $\sum_{k=1}^K N_k=N$. The fraction $p_k$ of elements in
the $k \mbox{th}$ cluster is $p_k=N_k/N$. The evolution of the $k\mbox{th}$ cluster
may be described by a variable $\chi_t(k)$ which gives the common state of the $N_k$
elements belonging to this cluster at time $t$. The periodic orbit adopted by the
state $\chi_t(k)$ of the $k \mbox{th}$ cluster can be expressed as a sequence of $P$
values $[\chi_1(k),\chi_2(k),\ldots,\chi_P(k)]$. The specific partition
$\{p_1,p_2,\ldots,p_K\}$ into $K$ clusters and the specific values taken by the
periodic orbit of each cluster depend on initial conditions and parameters of the
system.

When a GCM system falls into $K$ periodic clusters, the coupling function $H$ also
shows a periodic motion. As an illustration of this behavior, Fig. 1(a) shows a
typical situation in which a GCM system, with global coupling function $H=\bar{x}$,
displays two clusters, each in period two. In this case, the coupling function follows
a period-two motion.

Collective states consisting of two clusters have recently been observed in
an experimental array of globally interacting chemical elements \cite{Kiss}. This
clustered collective state is interesting to analyze in globally coupled maps. In this
simple situation, there is a fraction $p$ of elements in one cluster and a fraction
$(1-p)$ in the other cluster. Thus the global coupling functions from
Eqs.(\ref{coupling1})-(\ref{coupling4}) simplify to:

\begin{eqnarray}
\label{cluster1}
 H= & \langle x \rangle & =   p\chi_t(1)+(1-p)\chi_t(2), \\
 H= & \langle f \rangle & =   p f(\chi_t(2))+(1-p)f(\chi_t(2)),  \\
  H= & \Delta x & =  \sqrt{p(1-p)(\chi_t(2)-\chi_t(1))^2},  \\
  \label{cluster2}
   H= & \bar{x} & =  |\chi_t(1)|^{p}|\chi_t(2)|^{(1-p)} .
\end{eqnarray}
When the GCM system reaches a two-cluster state, the dynamics of the system reduces to
the two coupled maps
\begin{eqnarray}
\label{map1}
 \chi_{t+1}(1)& = & (1-\epsilon) f(\chi_t(1))+\epsilon H, \\
\label{map2}
 \chi_{t+1}(2)& = & (1-\epsilon) f(\chi_t(2))+\epsilon H;
\end{eqnarray}
where $H=H(\chi_t(1),\chi_t(2),p)$ is the reduced, two-cluster coupling function, as
in Eqs.(\ref{cluster1})-(\ref{cluster2}). If both cluster states $\chi_t(1)$ and
$\chi_t(2)$ fall in period-two orbits, the coupling function $H$ follows, in general,
a period-two motion, as shown in Fig. 1(a), although $H$ may become constant in some
circumstances (see Fig. 1(b) and Sec. IV). Let $H_1$ and $H_2$ be the values adopted
alternatively in time by $H$ in its period-two orbit for a given partition
$\{p,1-p\}$, as indicated on Fig. 1(a). The values $H_1$ and $H_2$ depend on i) the
functional form of $H$; ii) the parameters of the GCM system Eq.(\ref{gcm}); and iii)
the fraction $p$.

Consider then several GCM systems with the same parameters $r=1.7$ and $\epsilon =0.2$
but with different coupling functions such as those in Eqs.
(\ref{coupling1})-(\ref{coupling4}). Two clusters in period two can emerge in each of
these systems for some range of the fraction $p$. The resulting asymptotic orbits
$[H_1,H_2]$ of the respective coupling functions are shown in Fig. 2(a) as $p$ varies,
giving rise to a curve in each case. Note that each function $H$ possesses period-two
orbits only for a limited range of the fraction $p$. Figure 2(b) is a magnification of
Fig. 2(a) which shows that the dynamics of the backward and the mean field coupling
functions become equal for $p=0.5$; i.e., when the two clusters have equal sizes. In
this case, both coupling functions reach the constant value $H_1=H_2=0.365$. Notice
that the dispersion coupling function, $H=\Delta x$, only displays states with
$H_1=H_2$; that is, even when the two clusters in period two may have different sizes,
this particular global coupling always reaches a stationary value. The curves for the
other coupling functions are symmetrical with respect to the diagonal in Fig. 2(b),
which they cross for $p=0.5$. On the diagonal, the coupling functions are constant and
the two clusters evolve out of phase with respect to each other (Sec. IV).

Note also that the different global coupling functions perform a period-two motion
only on a restricted region of the plane $(H_1,H_2)$. It will be shown in Sec. III
that period-two orbits of any permutable coupling function will fall within the dashed
contour in Fig. 2.

In general, a coupling function $H$ of a GCM system in a collective state of two
clusters can reach various asymptotic periodic orbits for appropriate initial
conditions. Each Fig. 3(a) to 3(d) shows the regions on the space of parameters
$(p,\epsilon)$ for which a coupling function $H$ of a GCM in a two-cluster state
displays different periodic motions. The local parameter is fixed at $r=2$. Figures
3(a) and 3(b) correspond to the backward and forward mean field coupling,
respectively. Note the very different distributions of periodic regions for the
coupling functions in Figs. 3(a) and 3(b). It should be noticed that, besides the
collective periodic states for two clusters shown in Figs. 3(a)-3(d), there can exist
other states in a GCM system consisting of more than two periodic clusters for the
same values of the parameters $r$ and $\epsilon$, but corresponding to different
initial conditions.

The inverse problem of determining the global coupling function in
experimental systems is relevant since in general the specific functional
form of the acting coupling is not known. This can be a complicated problem
because, in addition, the exact form of the local dynamics may not
be extracted in most situations. However, if the local dynamics is known
some insight on the function $H$ of a globally coupled system can be
gained within the framework presented here.
For example, in the case of a dynamical system showing two period-two
clusters with partition $p$, the resulting asymptotic orbit $[H_1,H_2]$
can be obtained by measuring the cluster orbits and using Eqs. (\ref{map1}) and (\ref{map2}).
For different realizations of partition $p$, the curve $[H_1,H_2]$ can be drawn
as a function of $p$ on the plane $(H_1,H_2)$, and compared
with curves $[H_1,H_2]$ corresponding to known functional forms $H$
such as those in Fig.~2.

\section{Driven map analogy}
As shown in Ref. \cite {PC}, the clustering behavior of GCM systems can be analyzed
through its analogy with the dynamics of a single map subjected to an external drive.
In order to interpret the results of the preceding section, let us consider an
associated driven map
\begin{equation}
\label{driven}
 s_{t+1}=(1-\epsilon)f(s_t)+\epsilon L_t ,
\end{equation}
where $s_t$ is the state of the map at discrete time $t$, $f(s_t)$ is the same local
dynamics as in Eq.(\ref{gcm}), and $L_t$ is an external driving term assumed to be
periodic with period $P$. We denote the sequence of $P$ values adopted by the periodic
drive $L_t$ by $[L_1,L_2,\ldots,L_P]$. The analogy between a GCM system and a driven
map arises because in the former system (Eq. (\ref{gcm})) all the elements are
affected by the global coupling function in exactly the same way at all times, and
therefore the behavior of any element $x_t(i)$ in the GCM is equivalent to the
behavior of a single driven map (Eq. (\ref{driven})) with $L_t=H$ and initial
condition $s_o=x_o(i)$. Additionally, if a GCM system reaches a clustered, periodic
collective state, its corresponding coupling function $H$ follows in general a
periodic motion. Thus the associated driven map (Eq. (\ref{driven})) with a periodic
drive $L_t$ should display a behavior similar to that of an element belonging to a
periodic cluster in the GCM system. In particular, periodic drives resulting in
periodic orbits of $s_t$ in Eq. (\ref{driven}) may be employed to predict the
emergence of clustered, periodic states in a GCM (Eq. (\ref{gcm})), regardless of the
specific functional form of the global coupling $H$ and without doing direct
simulations on the entire GCM system.

The driven map is multistable; i.e., there can exist several attractors for the same
parameter values $r$ and $\epsilon$. Specifically, for a given periodic drive
$[L_1,L_2,\ldots,L_P]$, the map $s_t$ may reach a number of distinct asymptotic
responses $\bar{s}_t(j)$;  $(j=1,2, \ldots, J)$, all with the same period, and
depending on initial conditions. The orbits of $s_t$ with period $M$ can be expressed
as a sequence of values $[\bar{s}_1(j),\bar{s}_2(j),\ldots,\bar{s}_M(j)]$. The
correspondence between a GCM system (Eq. (\ref{gcm})) in a state of $K$ clusters with
period $P$ and its associated driven map (Eq. (\ref{driven})) can be established when
$M=P$ and $J=K$.

Using this analogy, the main features in Fig. 2 can now be explained. In terms of a
driven map subjected to a period-two drive $L_t=[L_1,L_2]$ and the same parameters
$r=1.7$ and $\epsilon=0.2$ as in the GCM systems considered in Fig. 2, the bounded
region on that figure contains the values of $L_1$ and $L_2$ for which the driven map
$s_t$ just possesses two distinct asymptotic orbits $(J=2)$ of period two $(M=2)$,
denoted by $[\bar{s}_1(1),\bar{s}_2(1)]$ and $[\bar{s}_1(2),\bar{s}_2(2)]$. For values
of $L_1$ and $L_2$ outside this diamond shaped region, the map $s_t$ may also reach a
number of asymptotic periodic orbits but none with both $J=2$ or $M=2$, at least on
the interval $-1 \leq L_1,L_2 \leq 1$. Because of the analogy drawn above, all
period-two motions $[H_1,H_2]$ of permutable coupling functions in GCM systems given
by Eq. (\ref{gcm}) with $r=1.7, \epsilon=0.2$ and displaying two clusters in period
two will fall on this bounded region of the plane $(H_1,H_2)$. Equivalently, a
collective state of two clusters in period two can emerge in a GCM system only if its
global coupling function has an orbit $[H_1,H_2]$ with values $H_1=L_1$ and $H_2=L_2$
contained within the bounded region of Fig. 2. The boundaries of that region vary as
the parameters $r$ and $\epsilon$ are changed. It is out of the scope of the present
work to establish how the bounded region for two clusters in period two, as well as
other regions for different clustered, periodic states, depend on parameters. However,
it is worth noticing that this dependence can serve to characterize GCM systems with
permutable coupling functions, and that this characterization can be obtained by the
sole use of an associated driven map.

The analogy between a GCM system with a given coupling function $H$ in a two-cluster
state and an associated driven map can be carried further by defining an associated
coupling function \cite{PC}
\begin{equation}
\Theta_H=H\left( \bar{s}_t(1),\bar{s}_t(2), p \right);
\end{equation}
that is, $\Theta_H$ is similar to the reduced two-cluster coupling function, such as
Eqs.(\ref{cluster1})-(\ref{cluster2}) but with the arguments $\chi_t(1)$ and
$\chi_t(2)$ from the cluster trajectories replaced by the driven map orbits
$\bar{s}_t(1)$ and $\bar{s}_t(2)$. The associated coupling function links the dynamics
of clusters in a GCM system to the dynamics of a single associated driven map. The
function $\Theta_H$ depends on the functional form of the coupling function $H$, on
the partition $\{p,1-p\}$ among the two clusters in the GCM, and on the orbits
$\bar{s}_t(1)$ and $\bar{s}_t(2)$, which themselves are function of the period-two
drive $[L_1,L_2]$ and the parameters $r$ and $\epsilon$. Thus for fixed $r$ and
$\epsilon$, and a given $H$, we have $\Theta_H=\Theta_H(L_1,L_2,p)$. The main point is
that, an equivalence between a GCM system Eq. (\ref{gcm}) in a two-cluster, period-two
state, and an associated driven map, Eq. (\ref{driven}), with a period-two drive
occurs when the following conditions are fulfilled
\begin{eqnarray}
\label{teta1}
 \Theta_H(\bar{s}_1(1),\bar{s}_1(2),p) & = & L_1, \\
\label{teta2}
 \Theta_H(\bar{s}_2(1),\bar{s}_2(2),p) & = & L_2.
\end{eqnarray}
Eqs. (\ref{teta1})-(\ref{teta2}) constitute a set of two nonlinear equations for $L_1$
and $L_2$, for a given $p$. The solution $[L_1^*,L_2^*]$ of Eqs.
(\ref{teta1})-(\ref{teta2}) predicts that the GCM possesses a state characterized by
the coupling function motion $[H_1=L_1^*,H_2=L_2^*]$ and cluster orbits
$[\chi_1(1),\chi_2(1)]=[\bar{s}_1(1),\bar{s}_2(1)]$,
$[\chi_1(2),\chi_2(2)]=[\bar{s}_1(2),\bar{s}_2(2)]$. The succession of solutions
$[L_1^*,L_2^*]$, as $p$ varies, yields the curve corresponding to a given $H$ in Fig.
2. Thus, each curve on the plane $(H_1,H_2)$ is parameterized by the fraction $p$.
Moreover, there exist solutions $[L_1^*,L_2^*]$ to Eqs. (\ref{teta1})-(\ref{teta2})
only for an interval of $p$. Therefore, the curves in Fig. 2 can, in principle, be
calculated {\it a priori} by using an associated driven map and just the specific
functional form of $H$ in each case. The range of possible cluster sizes, described by
the values of the fraction $p$ for which exist solutions to Eqs.
(\ref{teta1})-(\ref{teta2}), can also be predicted by this method. Similarly, the
regions of period two in Figs. 3(a)-3(d) can be obtained by varying the parameter
$\epsilon$ and calculating the interval of $p$ for which Eqs.
(\ref{teta1})-(\ref{teta2}) have solutions.

\section{Prediction and stability of clusters in systems with constant global coupling}

Another simple clustered collective state in GCM systems occurs when the coupling
function $H$ remains constant in time, i.e., $H=C$. This behavior may take place in a
GCM system  with a permutable coupling function when $K$ clusters, each having $N/K$
elements and period $K$, are evolving with shifted phases in order to yield a constant
value for $H$. That is, if the periodic orbits of $K$ identical-size clusters are
cyclically permuting in time, the resulting $H$ becomes constant. For those collective
states, the behavior of any of such clusters in the GCM system can be emulated by an
associated driven map subjected to a constant forcing $L_t=C$ \cite{PC2}. In the case
of a GCM displaying two equal size clusters in period two, this situation corresponds
to the intersection of $H$ with the diagonal in Fig. 2.  The cluster orbits are then
related as $\chi_t(1)=[\chi_1(1),\chi_2(1)]=[a,b]$ and
$\chi_t(2)=[\chi_1(2),\chi_2(2)]=[b,a]$. On the other hand, the associated driven map
with $L_t=C$ has a unique asymptotic period-two orbit $\bar{s}_t=[\alpha,\beta]$ on a
range of $C$, where $\alpha$ and $\beta$ are functions of $C$. The associated coupling
function $\Theta_H$ also simplifies in such case. For the reduced, two-cluster
couplings in Eqs (\ref{cluster1})-(\ref{cluster2}), the corresponding associated
coupling functions become
\begin{eqnarray}
 \label{rteta1}
\Theta_{\langle x \rangle}(\alpha,\beta) & = & \frac{1}{2}(\alpha+\beta) \\
 \Theta_{\langle f \rangle}(\alpha,\beta) & = & \frac{1}{2}[f(\alpha)+f(\beta)] \\
 \Theta_{\Delta x}(\alpha,\beta) & = & \frac{1}{2}|\alpha-\beta| \\
  \label{rteta2}
 \Theta_{\bar{x}}(\alpha,\beta) & = & |\alpha|^{1/2}|\beta|^{1/2}.
 \end{eqnarray}
Then Eqs. (\ref{teta1})-(\ref{teta2}) with $L_1=L_2=C$ reduce to the single equation
\begin{equation}
\label{redteta}
 \Theta_H(\alpha,\beta)=C,
\end{equation}
which can be seen as an equation for $C$, for given values of the parameters
$\epsilon$ and $r$. The solution $C=C_*$ of Eq. (\ref{redteta}) provides a complete
description of the GCM state since then $a=\alpha(C_*), b=\beta(C_*)$ and $H=C_*$.
Figure 4 shows the bifurcation diagram of $s_t$, Eq. (\ref{driven}), as a function of
the constant drive $L_t=C$ up to period two, with fixed parameters $r=2$ and
$\epsilon=0.4$. The fixed point region in this diagram corresponds to one stationary
cluster (i.e., a synchronized collective state) in the GCM, Eq. (\ref{gcm}), with
constant $H$. The period-two window corresponds to the values $\alpha$ and $\beta$
adopted by the driven map on this range of $C$. Once $\alpha(C)$ and $\beta(C)$ are
known from the bifurcation diagram, the function $\Theta_H(C)$ associated to any
global coupling function in a GCM can be readily obtained, assuming that the GCM is in
a state of two equal size clusters $(p=0.5)$, evolving out of phase with respect to
each other. In Figure 4, the $\Theta_H$ functions associated to the four global
couplings in Eqs.(\ref{coupling1})-(\ref{coupling4}) with fixed $\epsilon$ are shown
as function of $C$. As stated above, the solutions $C_*$ to Eq. (\ref{redteta})
correspond to states in GCM systems with a coupling function reaching a stationary
value $H=C_*$. Thus, the intersections of the $\Theta_H$ curves with the diagonal in
Fig. 4 give all the possible states of GCM systems that maintain a constant $H$,
either with one stationary cluster (if the intersection occurs on the fixed point
window of the bifurcation diagram of $s_t$) or with two clusters in period two (if the
intersection occurs on the period two window of the diagram). Note that both the
backward and the forward mean field couplings have the same two-cluster, period-two
solution $\langle x \rangle =\langle f \rangle = C_*=0.416$, but these couplings have
different functional dependence on the constant drive $L_t=C$. The coincidence of the
couplings $\langle x \rangle$ and $\langle f \rangle$ for $p=0.5$ was already seen in
Fig. 2(b). Similarly, the geometric mean coupling $H=\bar{x}$ gives only one
two-cluster, period-two solution at $\bar{x}=C_*=0.305$. In contrast, the dispersion
global coupling, $H=\Delta x$, has three intersections with the diagonal: one
corresponds to the synchronized stationary state in the associated GCM, with $\Delta
x= C_*=0$, and the other two correspond to different clustered states of the GCM, each
consisting of two equal size clusters in period two, with $\Delta x=C_*=0.083$ and
$\Delta x=C_*=0.25$, respectively. All the states predicted by the intersections of
the different $\Theta_H$ with the diagonal in Fig. 4, except one, are readily found in
simulations on the corresponding GCM systems for appropriated initial conditions in
each case. Actually, for a GCM with the coupling $H=\Delta x$, the predicted
two-cluster, period-two state with $\Delta x=C_*=0.083$ is unstable: it is never
achieved in simulations on the GCM system, even when the initial conditions are chosen
very close to that state. What is observed, instead, is the evolution of the GCM
system towards either the stationary one cluster state with $\Delta x=0$ or the
two-cluster, period-two state with $\Delta
 x=0.25$. Thus, in addition to being predicted by the solutions of the equation
$\Theta_H=C$, the observed clustered states of a GCM displaying constant coupling must
be stable, which implies some stability condition on the solutions. It can be shown
(see Appendix) that for coupling functions satisfying $\sum_{i=1}^N
\partial H /
\partial x_i = 0$, the condition $d \Theta_H/d C > 1$ at the intersection with the
diagonal implies that the corresponding solution is unstable. This is the case of the
global coupling $H=\Delta x$. Note that the solution at $C_*=0.083$ is the only one
for which $\left. \frac{d \Theta}{d C}\right|_{C_*} > 1$ in Fig. 4 and therefore it
is unstable, independently of the cluster fraction $p$. For the coupling $H= \langle f
\rangle$, a stability analysis of  states consisting of two or three clusters in
period three
 has been performed by Shimada and Kikuchi \cite{Tokuzo}. However,
the  simple criterium for instability $\left. \frac{d \Theta}{d C}\right|_{C_*} > 1$
does not apply for $H= \langle f \rangle$.

Constant coupling functions may also occur in GCM systems with different cluster
sizes; that is the case of a GCM possessing dispersion global coupling $H=\Delta x$
and displaying two clusters with any partition $\{p,1-p\}$, as seen in Fig. 1(b).
Figure 5 shows the associated function $\Theta_{\Delta x}$ with fixed $\epsilon$ as a
function of the constant drive $C$ for several values of the fraction $p$. There exist
a critical fraction $p_c=0.25$ bellow which only one solution corresponding to one
stationary cluster, i.e., synchronization, can appear in the GCM system. Above this
critical fraction, two states, each consisting of two clusters in period two, are
additionally predicted by the solutions $C=C_*$ of $\Theta_{\Delta x}=C$. These
solutions emerge as a pair: one solution is always unstable since $d\Theta_H/dC
|_{C_*} > 1$ there, and the other is the one two-cluster, period-two collective state
with constant $\Delta x$ that is actually observed in simulations on the corresponding
GCM system. For the fraction $p_c=0.25$, there is a two-cluster, period-two solution
$\Theta_{\Delta x}=C_*=0.125$ that is marginally stable.

\section{Conclusions}

Most studies on GCM lattices and other globally coupled systems have assumed mean
field coupling. However, other forms of global coupling may be relevant in some
situations. We have analyzed, in a general framework, the clustering behavior in GCM
systems subjected to permutable global coupling functions by considering the dynamics
of the coupling functions.

We have shown that different GCM systems can be represented
by the orbits of their coupling functions on a common space.
For simplicity, only collective states in GCM
systems consisting of two clusters in period two were considered. We have shown that
the functional form of the global coupling in a GCM system determines the periodicity
of its motion and the possible distributions of elements among the clusters.
The existence of
a well defined interval of possible partitions among two clusters,
out of which no clusters emerge in the system,
has been observed experimentally \cite{Kiss}.
In experimental or natural situations where clustering occurs, the specific
functional form of the coupling is in general unknown.
The present study may be useful to obtain insight into the acting global coupling
function in practical situations.

We have employed a previously introduced analogy between a GCM system and a single
externally driven map \cite{PC} in order to give a unified interpretation of the observed
clustering behavior of the GCM systems considered in this article. A periodically
driven map with local periodic windows can display multiple asymptotic periodic
responses which are similar to cluster orbits in a GCM system with permutable $H$.
This analogy implies that dynamical clustering can occur in any GCM system with a
permutable coupling function and periodic windows in the local dynamics. The presence
of windows of stable periodic orbits in the local map is essential for the emergence
of clusters.
In fact, no clustering is observed in a GCM system if the
local maps do not have periodic windows \cite{Gallego}; what is observed instead is
synchronization or nontrivial collective behavior, i.e., an ordered temporal evolution
of statistical quantities coexisting with local chaos.

The associated coupling function derived from
the driven map analogy is particularly simple to use
in the prediction of clustered states in GCM systems with two
equal size clusters and exhibiting constant global coupling $H=C$. The associated
coupling function $\Theta_H$ can be directly constructed from the bifurcation diagram
of the steadily driven map. The cluster states are obtained from the solutions of Eq.
(\ref{redteta}) and can be represented graphically in a simple way.  Although Eq.
(\ref{redteta}) has been used for the case of two clusters in period two, it can also
be applied to find GCM states consisting of $K$ equal size clusters in period $K$. In
addition, the associated coupling function $\Theta_H$ carries
information about the stability of the predicted two-cluster states.
In particular, for the family of coupling function satisfying property (\ref{even}),
the stability condition of  clustered states
 in a GCM with constant $H=C_*$ is directly given by the slope $ d\Theta_H/d C|_{C_*}$. The
example of a GCM system with dispersion coupling function $H=\Delta x$ reveals that a
constant coupling can also be maintained by clusters of different sizes. Our method
based on Eq. (\ref{redteta}) also predicts successfully the cluster states in these
situations.

The driven map analogy suggests that the emergence of clusters should be a common
phenomenon which can be expected in various dynamical systems formed by globally
interacting elements possessing stable periodic orbits on some parameter range of
their individual dynamics. The examples presented here show that
progress
in the understanding of the collective behavior of globally coupled
systems can be achieved
by investigating their relation to a driven oscillator.

\section*{Acknowledgment}
This work has been supported by Consejo de Desarrollo Cient\'{\i}fico,
Human\'{\i}stico y Tecnol\'ogico of the Universidad de Los Andes, M\'erida, Venezuela.

\appendix
\section*{On the relation between Stability and the Associated Coupled Function}
Consider a general GCM system with any global permutable coupling function. Suppose
that the system reaches a state consisting of two clusters. Then the dynamics of the
system reduces to two coupled maps Eqs. (\ref{map1})-(\ref{map2}), i.e.,
\begin{equation}
\begin{array}{lll}
\chi_{t+1}(1)= & (1-\epsilon) f(\chi_t(1))+\epsilon H(\chi_t(1),\chi_t(2),p)= &
F(\chi_t(1),\chi_t(2)),  \\

\chi_{t+1}(2)= & (1-\epsilon) f(\chi_t(2))+\epsilon H(\chi_t(1),\chi_t(2),p)= &
G(\chi_t(1),\chi_t(2)).
\end{array}
\end{equation}

If the two clusters are in period-two orbits, the stability of this collective state
in the GCM is given by the eigenvalues of the product of Jacobian matrices
\begin{equation}
\label{jacob}
 {\bf J}={\bf J}_1 {\bf J}_2=\prod_{i=1}^2 \left(
\begin{array}{cc}
  \frac{\partial F}{\partial {\chi_i(1)}} & \frac{\partial F}{\partial {\chi_i(2)}} \\
                                       &                                      \\
  \frac{\partial G}{\partial{ \chi_i(1)}} &\frac{\partial G}{\partial {\chi_i(2)}}
\end{array}
\right) \, .
\end{equation}

If the two clusters move out of phase, the asymptotic state of the GCM can be
described by the two orbits $\chi_t(1)\equiv [\chi_1(1),\chi_2(1)]= [a,b]$ and
$\chi_t(2) \equiv [\chi_1(2),\chi_2(2)]=[b,a]$, which satisfy
\begin{equation}
\begin{array}{ll}
  b =& (1-\epsilon)f(a)+\epsilon H(a,b,p), \\
  a =& (1-\epsilon)f(b)+\epsilon H(a,b,p).
\end{array}
\end{equation}

For the local dynamics $f(x)=1-rx^2$, one gets
\begin{equation}
{\bf J}_1 =  \left(
\begin{array}{cc}
  -2(1-\epsilon)r a+\epsilon H_a & \epsilon H_b \\
                                       &              \\
  \epsilon H_a & -2(1-\epsilon)r b+\epsilon H_b
\end{array}
\right) ,
\end{equation}
\begin{equation}
{\bf J}_2 = \left(
\begin{array}{cc}
  -2(1-\epsilon)r b+\epsilon H_b & \epsilon H_a \\
                                       &              \\
  \epsilon H_b & -2(1-\epsilon)r a+\epsilon H_a
\end{array}
\right) ,
\end{equation}
where
\begin{equation}
H_a \equiv \left.\frac{\partial H(\chi_t(1),\chi_t(2))}{\partial \chi_t(1)}\right|_{
 \chi_t(1)=a, \, \chi_t(2)=b}
 = \left.
\frac{\partial H(\chi_t(1),\chi_t(2))}{\partial \chi_t(2)}\right|_{\chi_t(1)=b, \,
\chi_t(2)=a} ,
\end{equation}
and
\begin{equation}
H_b \equiv \left.\frac{\partial H(\chi_t(1),\chi_t(2))}{\partial
\chi_t(2)}\right|_{\chi_t(1)=a, \, \chi_t(2)=b}= \left.\frac{\partial
H(\chi_t(1),\chi_t(2))}{\partial \chi_t(1)}\right|_{\chi_t(1)=b, \, \chi_t(2)=a} .
\end{equation}
Consider now the dispersion coupling function, Eq. (\ref{coupling3}). This coupling
belongs to the family of functions of $N$ variables
\begin{equation}
 H(x_1,x_2,\ldots,x_N)=\sum_{i=1}^N h_e(x_i-\langle x \rangle),
\end{equation}
where $h_e$ is any even function of its argument. It can be straightforwardly shown
that this family of functions possesses the property
\begin{equation}
\label{even}
\sum_{i=1}^N\frac{\partial H}{\partial x_i} = 0.
\end{equation}
Therefore, in a two cluster state, any $H$ in this family of global coupling functions
satisfies
\begin{equation}
\label{prop2}
 \frac{\partial H(\chi_t(1),\chi_t(2))}{\partial \chi_t(1)} + \frac{\partial
H(\chi_t(1),\chi_t(2))}{\partial \chi_t(2)}= 0.
\end{equation}
If the two clusters evolve out of phase with respect to each other, and additionally
the GCM has a coupling $H$ with property (\ref{even}), then the two eigenvalues of the
matrix ${\bf J}$ in Eq. (\ref{jacob}) become identical and their value is
\begin{equation}
\lambda= 2r\epsilon(\epsilon-1)(a H_b + b H_a)+4r^2 a b (\epsilon-1)^2.
\end{equation}
The stability criterion of this state is given by the modulus of the eigenvalue
$\lambda$; that is, $|\lambda|>1$ ($|\lambda|<1$) implies that the state is unstable
(stable). The values $a$ and $b$ are, respectively, the values of $\alpha$ and $\beta$
at the intersection of the function $\Theta_{\Delta x}$ with the diagonal in Fig. 4.

Let us analyze the relationship between the eigenvalue $\lambda$ and the derivative
$d\Theta_{\Delta x}/dC$ at the intersection points with the diagonal in Fig. 4 or Fig.
5. In general,
\begin{equation}
\frac{d\Theta}{d C}= \frac{\partial \Theta}{\partial \alpha}\frac{\partial
\alpha}{\partial C}+\frac{\partial \Theta}{\partial \beta}\frac {\partial
\beta}{\partial C} ,
\end{equation}
where
\begin{equation}
\label{sig1}
 \alpha=  \frac{-1+R}{2r(\epsilon-1)},
\end{equation}
\begin{equation}
\label{sig2}
 \beta=  \frac{-1-R}{2r(\epsilon-1)} ,
\end{equation}
and
\begin{equation}
R=\left(-3-4r\epsilon^2C+4r\epsilon C+ 4r-8r\epsilon+4r\epsilon^2\right)^{1/2}.
\end{equation}
Since
\begin{equation}
\frac{\partial \alpha}{\partial C}=-\frac{\partial \beta}{\partial C}=
\frac{1}{2r(\epsilon-1)}\frac{\partial R}{\partial C}= -\frac{\epsilon}{R} \, ,
\end{equation}
then
\begin{equation}
\frac{d\Theta}{d C}= -\frac{\epsilon}{R}\left(\frac{\partial \Theta}{\partial
\alpha}-\frac{\partial \Theta}{\partial \beta}\right).
\end{equation}
Since $\Theta_{\Delta x}$ has the same functional form as $H=\Delta x$, then
$\Theta_{\Delta x}$ also satisfies property (\ref{prop2}), that is
\begin{equation}
\frac{\partial \Theta_{\Delta x}}{\partial \alpha}=-\frac{\partial\Theta_{\Delta
x}}{\beta} ,
\end{equation}
and therefore
\begin{equation}
\frac{d \Theta_{\Delta x}}{ d C}= -\frac{2\epsilon}{R} \frac{\partial \Theta_{\Delta
x}}{\partial \alpha}.
\end{equation}
Let $C=C_*$ be a value of $C$ corresponding to the intersection of $\Theta_{\Delta x}$
with the diagonal in Fig. 4. Then $\alpha(C_*)=a$ and $\beta(C_*)=b$; and
$\Theta_{\Delta x}(C_*)=H(a,b)$, which gives
\begin{equation}
\label{dT}
 \left. \frac{d \Theta_{\Delta x}}{ d C} \right|_{C_*}=-\frac{2\epsilon}{R}
H_a.
\end{equation}
Using the fact that $H_a=-H_b$ from Eq. (\ref{prop2}), the eigenvalue $\lambda$
becomes
\begin{equation}
\label{l2}
 \lambda=2 r \epsilon(\epsilon-1)H_a(b-a)+4r^2ab(\epsilon-1).
\end{equation}
Eqs.(\ref{sig1}) and (\ref{sig2}) with $C=C_*$ give the values $a$ and $b$,
respectively. Then, substitution of these values and $H_a$ from Eq.(\ref{dT}) in Eq.
(\ref{l2}) yields
\begin{equation}
\lambda=R^2 \left( \left. \frac{d \Theta_{\Delta x}}{ d C} \right|_{C_*}-1\right)+1.
\end{equation}
Therefore, the condition $d\Theta/dC|_{C_*} > 1$ implies that $|\lambda| > 1$, and
thus the two-cluster, period-two solution with $C_*=0.07$ given by the intersection of
$\Theta_{\Delta x} $ with the diagonal in Fig. 4 is unstable. Similarly, the solutions
$C_*$ of $\Theta_{\Delta x}=C$ for the different curves in Fig. 5 for which
$d\Theta/dC|_{C_*} > 1$, are unstable.

Note that the above stability result for $H=\Delta x$ is also valid for any global
coupling function satisfying property (\ref{even}).

\begin{figure}
\epsfig{file=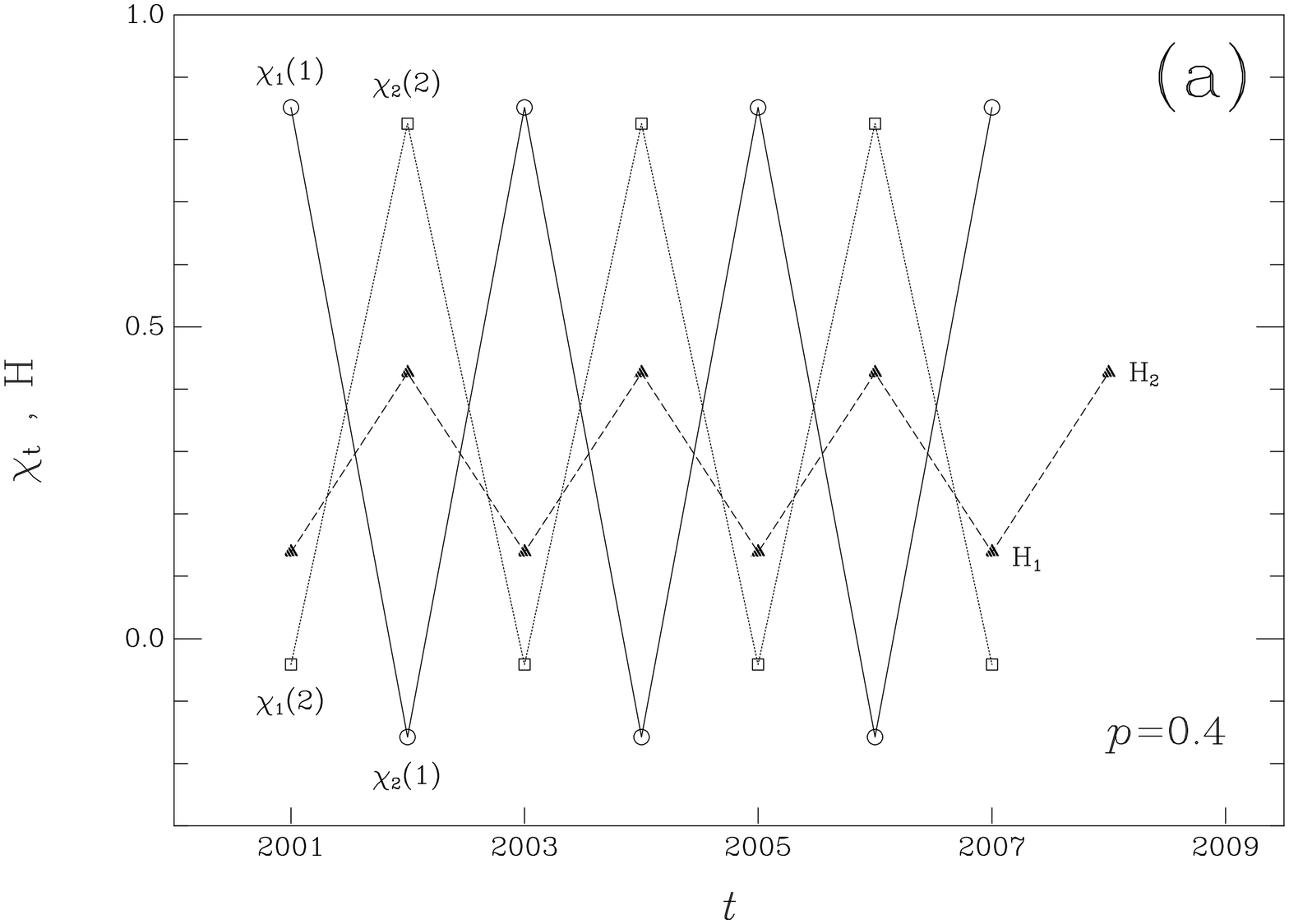, width=0.5\textwidth ,angle=270, clip=}
\epsfig{file=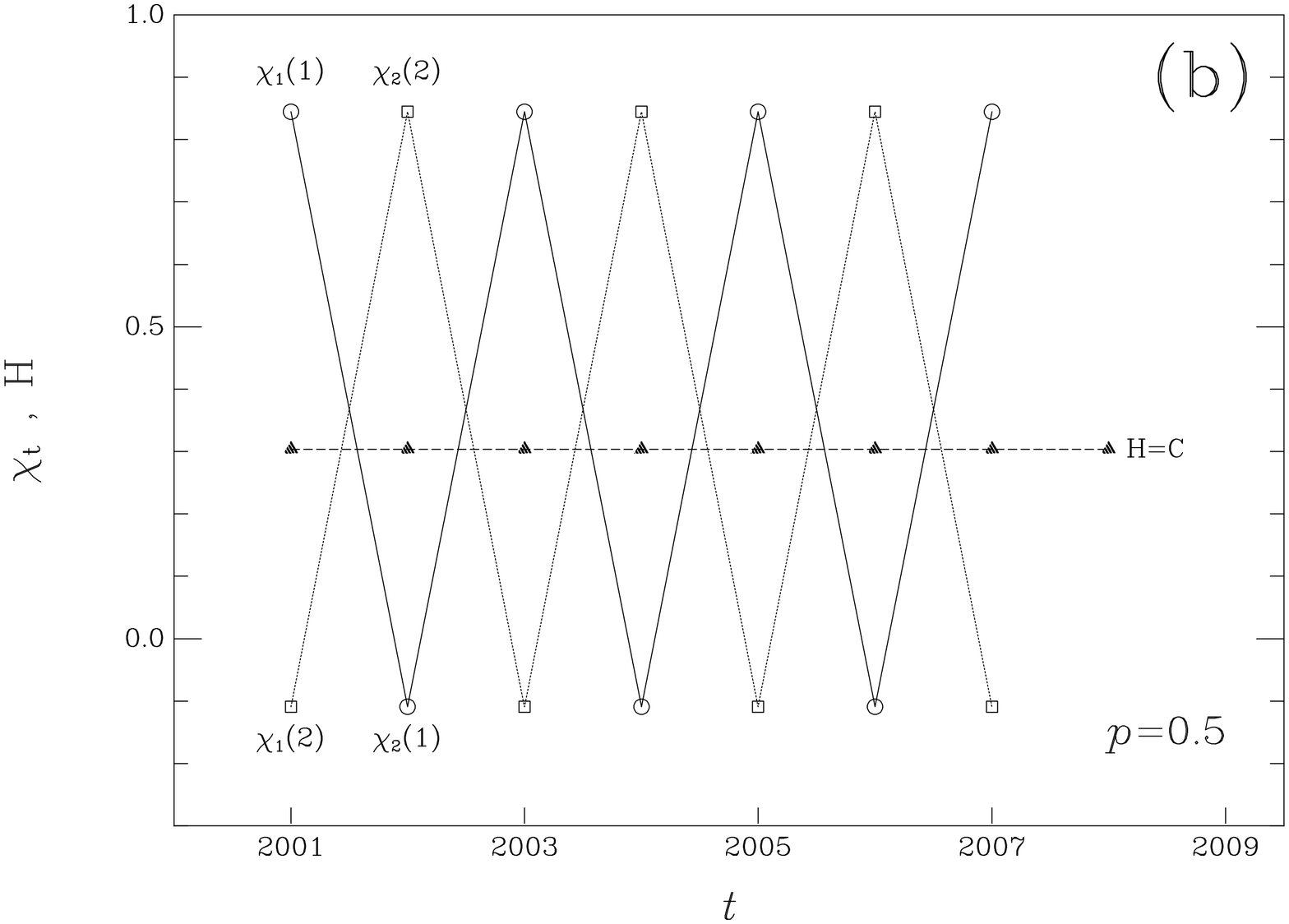, width=0.5\textwidth ,angle=270, clip=}
\caption{Dynamics of the coupling $H=\bar{x}$ (triangles) in a GCM system Eq.
(\ref{gcm}) with parameters $r=1.7$, $\epsilon=0.2$, displaying two clusters, each in
period two. Cluster orbits are $\chi_t(1)=[\chi_1(1),\chi_2(1)]$ (circles) and
$\chi_t(2)=[\chi_1(2),\chi_2(2)]$ (squares). (a) For the partition $p=p_1=0.4$ and
$p_2=0.6$, $H$ follows a period-two motion, adopting the values $[H_1,H_2]$. (b) For
$(p=0.5)$, the two clusters evolve out of phase with respect to each other and $H$
remains constant at the value $H=C=0.3037$.}
\end{figure}
\begin{figure}
\epsfig{file=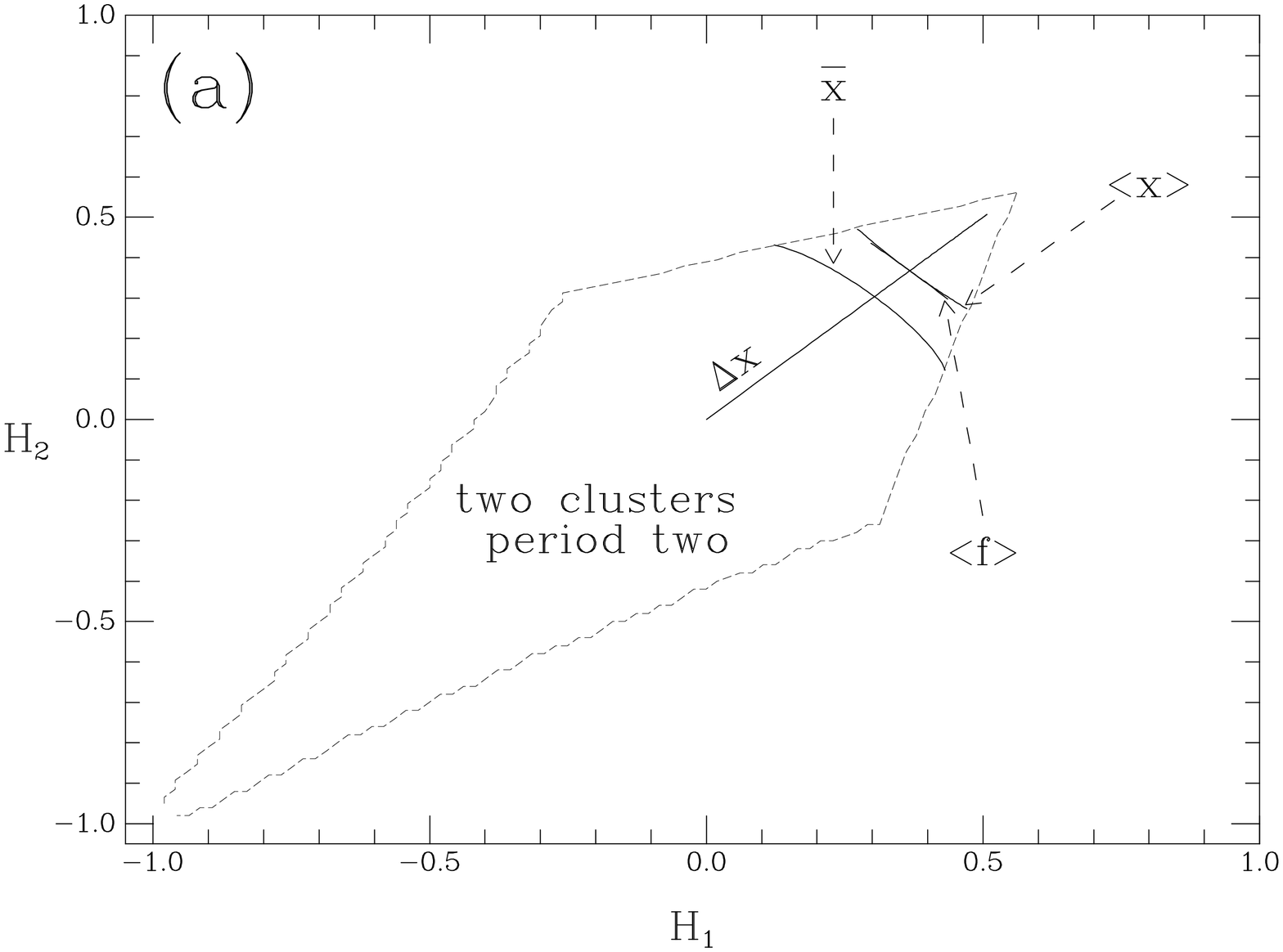, width=0.75\textwidth ,angle=270, clip=}
\epsfig{file=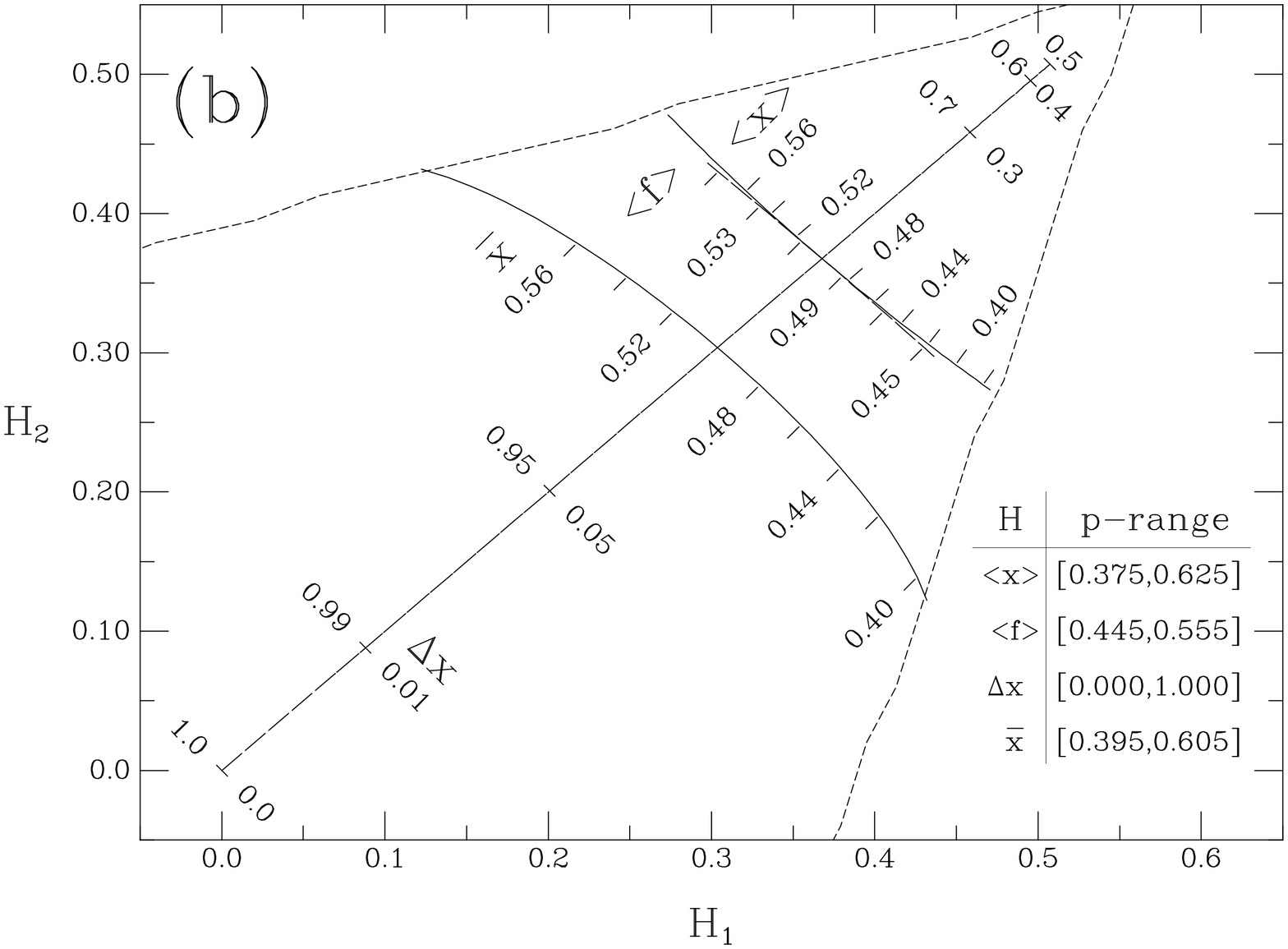, width=0.75\textwidth ,angle=270, clip=}
\caption{ a) Curves of period-two orbits $[H_1,H_2]$ on the plane $(H_1,H_2)$ as $p$
varies for the
 four coupling functions Eqs. (\ref{cluster1})-(\ref{cluster2}) in corresponding
  GCM systems displaying two clusters. Parameters are the same for the four
  systems, $r=1.7$, $\epsilon=0.2$.  The boundaries of the region where period-two orbits of any
permutable $H$ may take place is indicated with dashed lines. b) Magnification of a).
Labels identify the curve associated to each $H$ and the numbers besides the marks
along each curve indicate the corresponding values of the cluster fraction $p$. The
range of possible values of $p$ for the different curves is also displayed on the
figure.}
\end{figure}
\begin{figure}
\epsfig{file=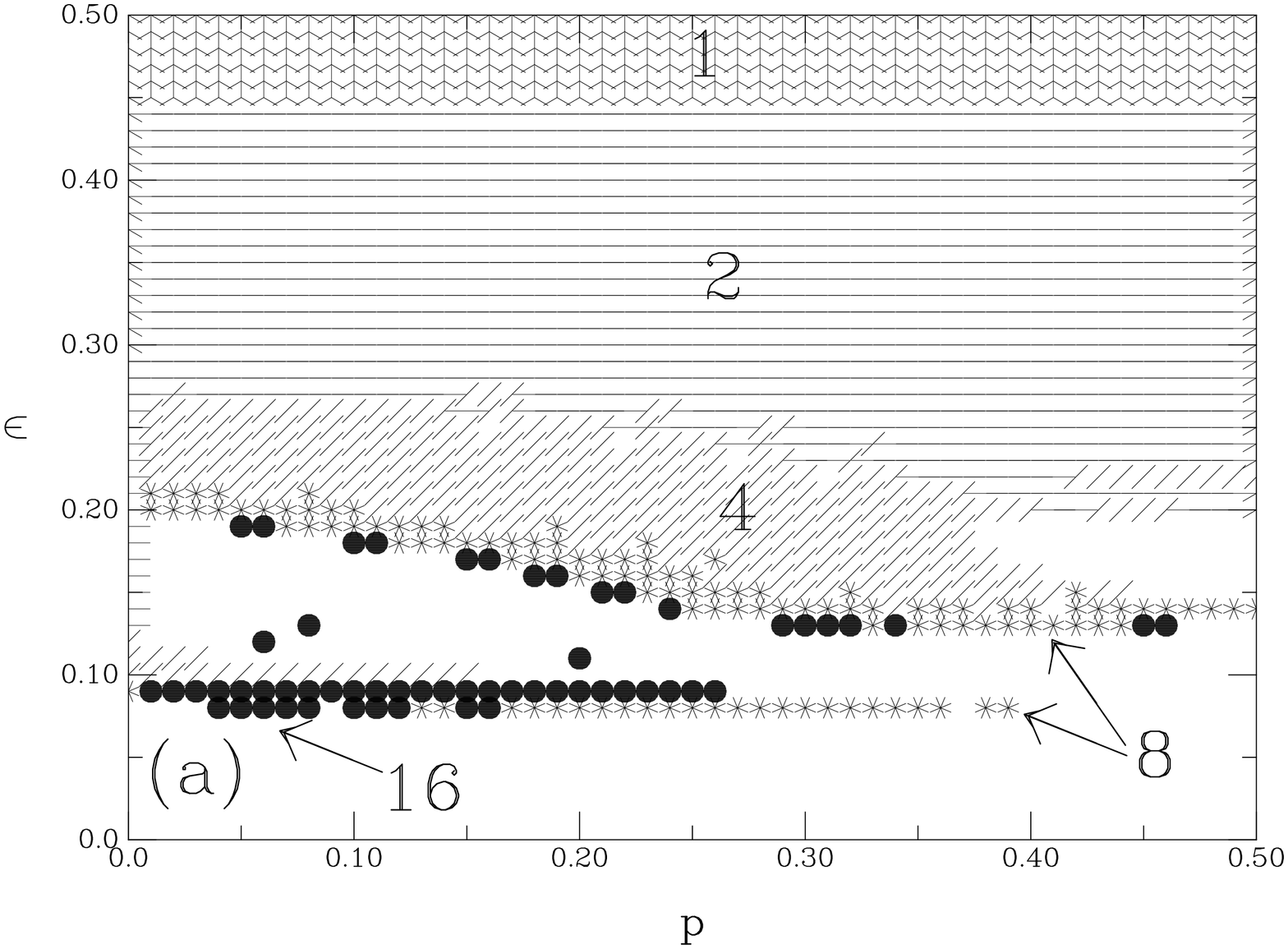, width=0.45\textwidth ,angle=270, clip=}
\epsfig{file=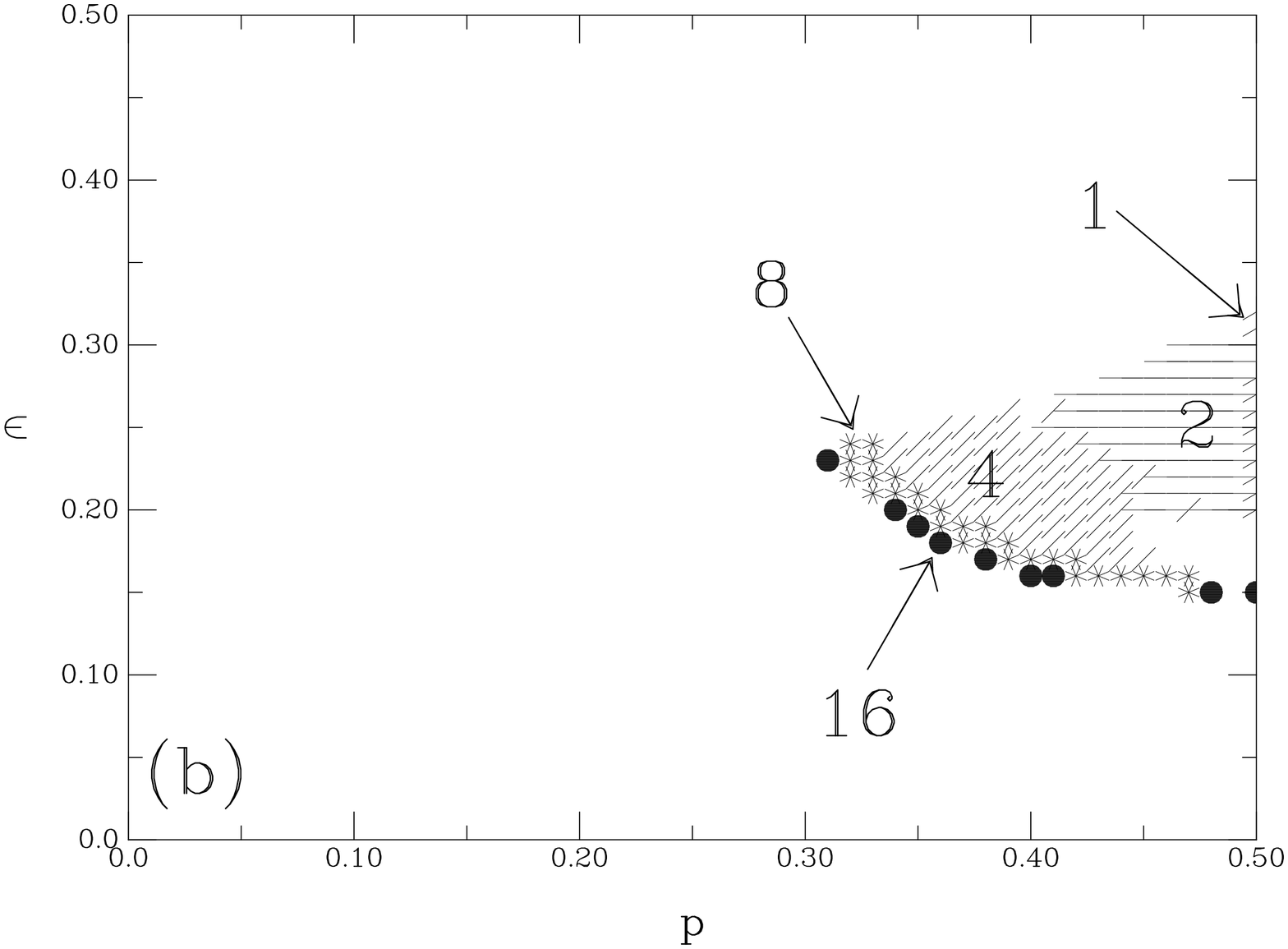, width=0.45\textwidth ,angle=270, clip=}
\epsfig{file=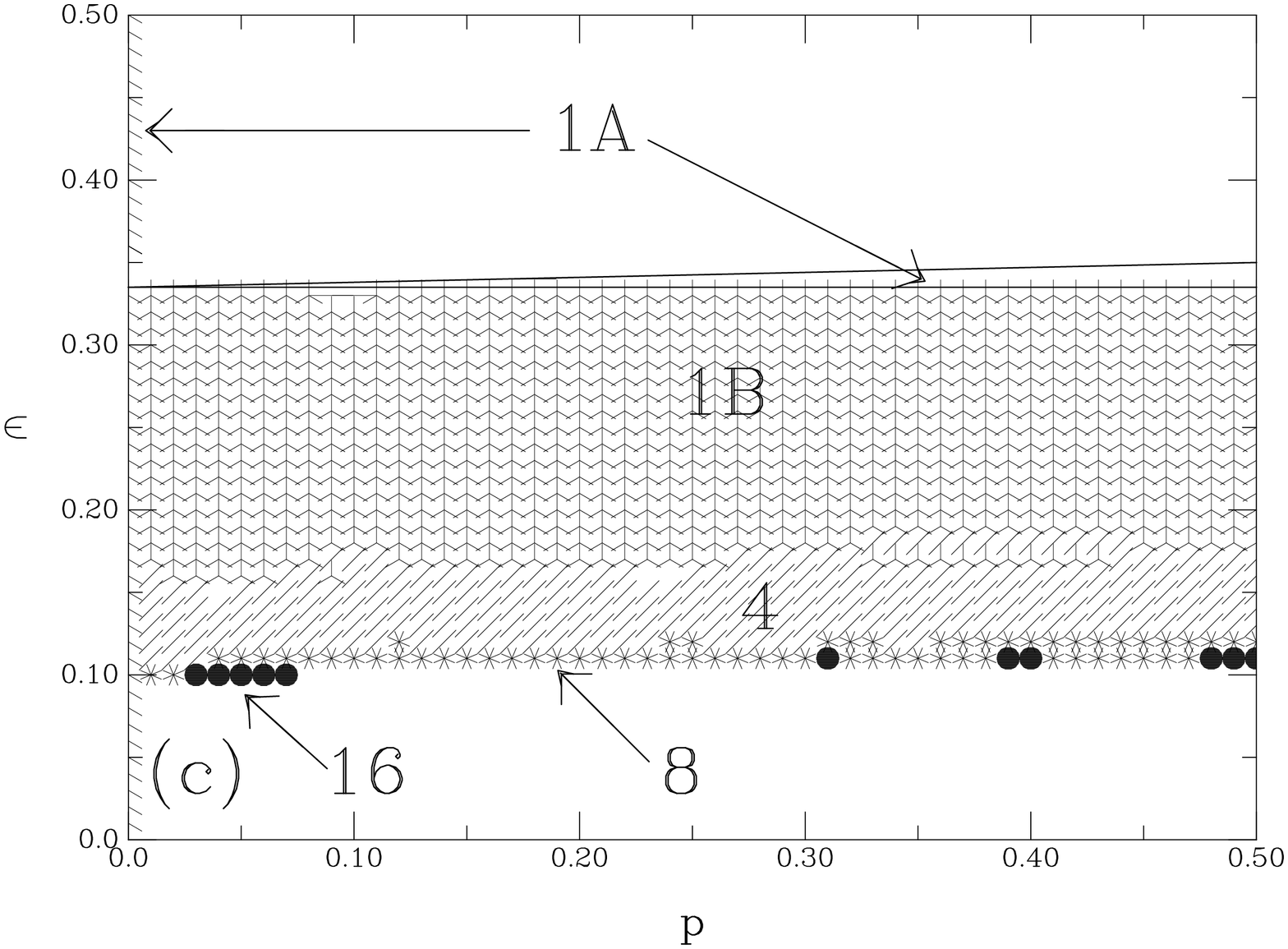, width=0.45\textwidth ,angle=270, clip=}
\epsfig{file=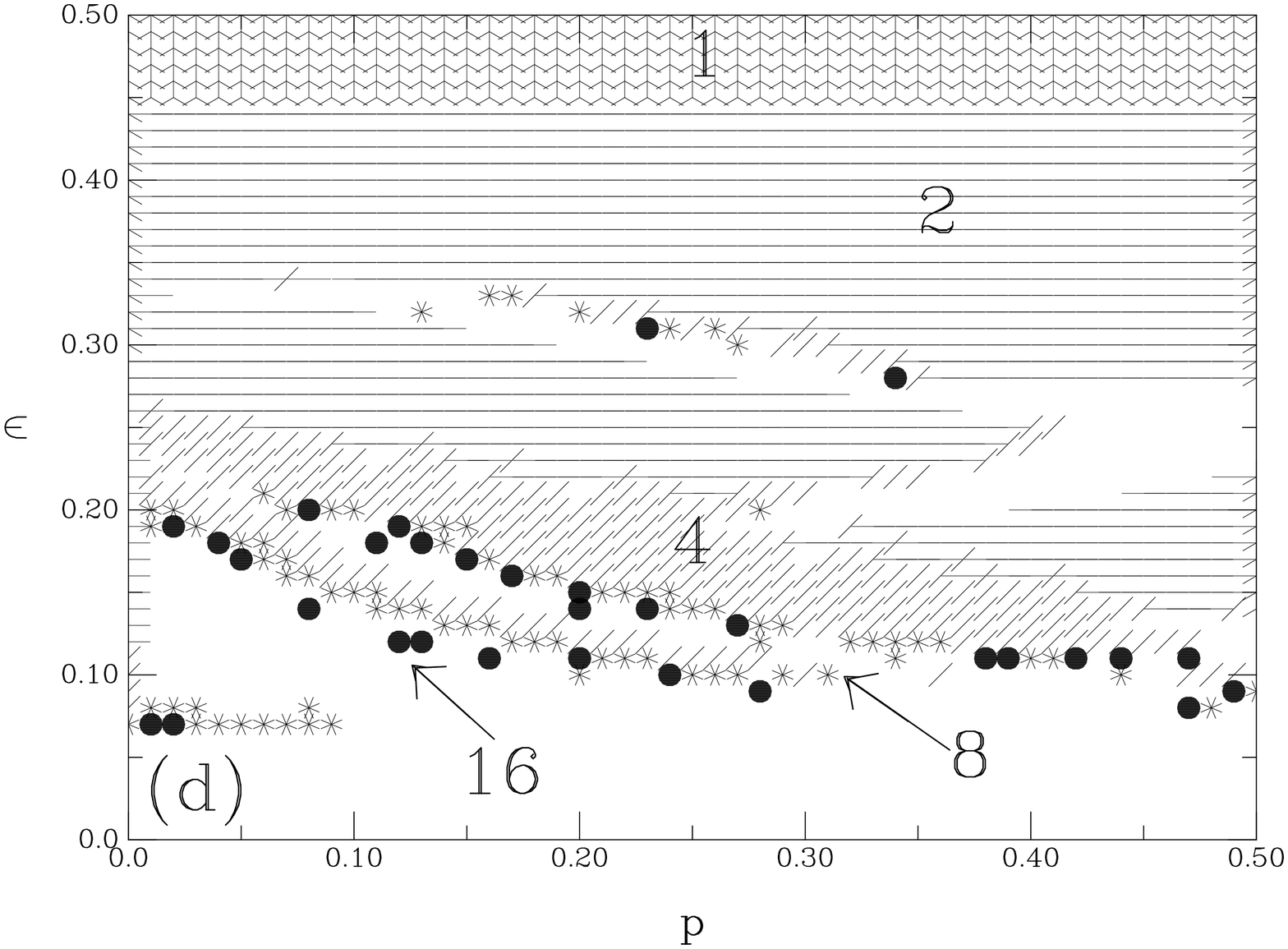, width=0.45\textwidth ,angle=270, clip=}
\caption{ Regions of periodic motions on the plane $(p,\epsilon)$ for different
couplings $H$ in GCM systems displaying two clusters. Local parameter is fixed at
$r=1.7$. The numbers on each region indicate the period of $H$ on that region.
 a) $H=\langle x \rangle$. b) $H=\langle f \rangle$. c) $H=\Delta x$; 1A: there is
 only one stationary cluster with constant $H=0$ along the line $p=0$,
 bistability occurs on the edge-shaped region: a state of one stationary cluster with
 $H=0$ coexist with a state of
 two out of phase clusters with constant, nonvanishing $H$; 1B: there are two out of
 phase clusters giving constant $H\neq 0$. d) $H=\bar{x}$. }
\end{figure}
\begin{figure}
\epsfig{file=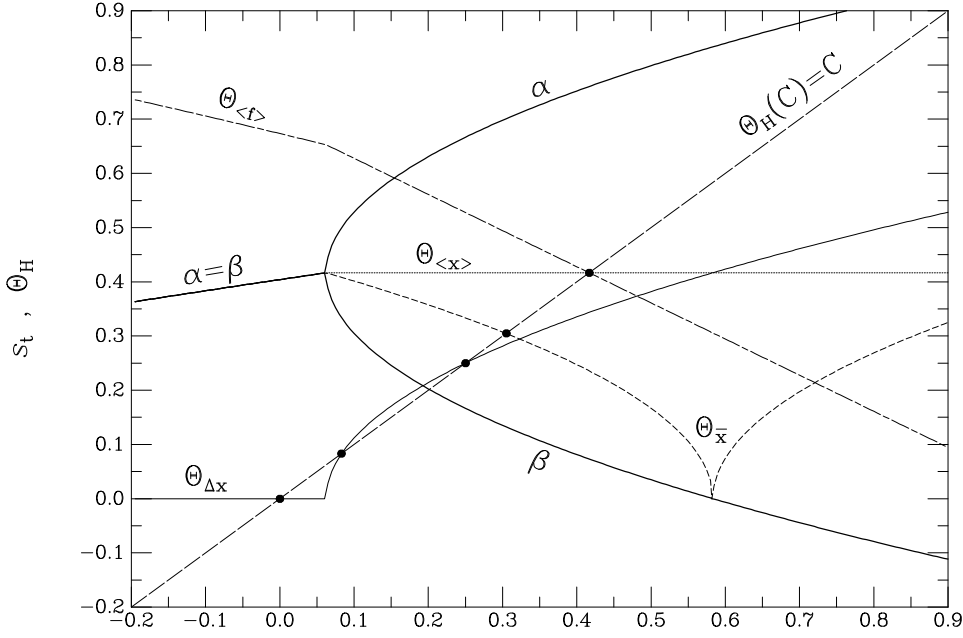, width=0.5\textwidth ,angle=270, clip=}
\caption{ Bifurcation diagram of the driven map $s_t$, Eq. (\ref{driven}), with
$L_t=C$, as a function of $C$. The asymptotic orbits of $s_t$ are drawn with solid
lines. The values $\alpha$ and $\beta$ on the period-two window of the driven map are
indicated; $\alpha=\beta$ on the fixed point window. The associated coupling functions
$\Theta_H$ from Eqs. (\ref{rteta1})-(\ref{rteta2}) are also shown vs. $C$. The
intersections with the diagonal $\Theta_H=C$ are indicated by black dots and they
correspond to the solutions $C_*$ of Eq. (\ref{redteta}) }.
\end{figure}
\begin{figure}
\epsfig{file=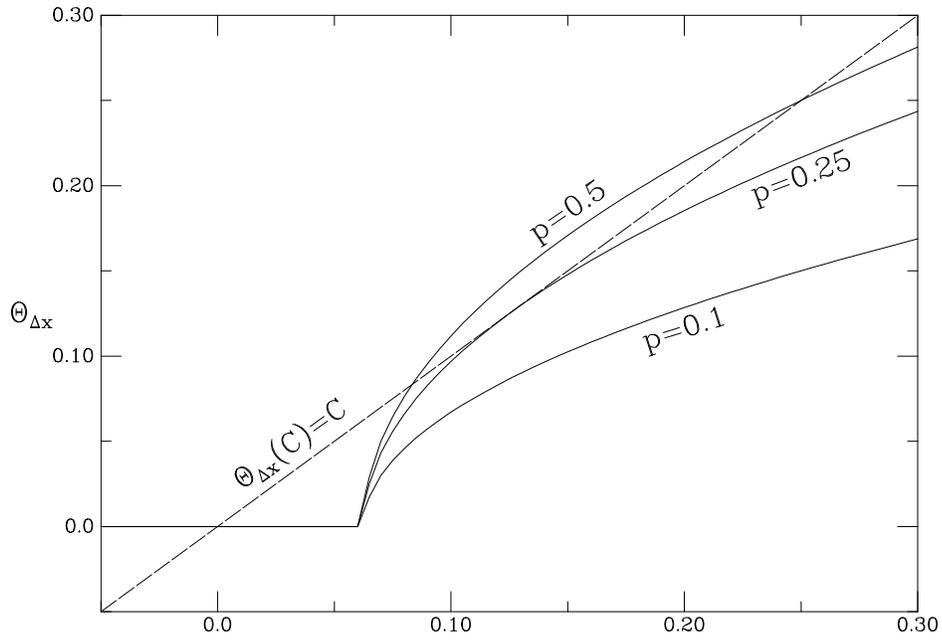, width=0.5\textwidth ,angle=270, clip=}
\caption{ The associated coupling function $\Theta_{\Delta x}$ vs. $C$ for different
values of the fraction $p$. The critical fraction is $p_c=0.25$. Intersections with
the diagonal give the solutions $C_*$. Solutions for which $\frac{d \Theta}{d
C}|_{C_*} > 1$ are unstable. }
\end{figure}

\end{document}